\documentclass[twocolumn,prl,aps,superscriptaddress]{revtex4-1}

\usepackage{blindtext} 

\usepackage[sc]{mathpazo} 
\usepackage[T1]{fontenc} 

\usepackage[english]{babel} 

\usepackage{enumitem} 
\setlist[itemize]{noitemsep} 

\usepackage{hyperref} 

\usepackage{xcolor}

\usepackage[normalem]{ulem}

\hypersetup{
    colorlinks=true,
    linkcolor=blue,
    filecolor=magenta,      
    urlcolor=blue,
}

\usepackage{braket}

\usepackage{graphicx}

\begin{document}
\title{Decoding the DC and optical conductivities of disordered MoS$_2$ films: an inverse problem} 
\author{F. R. Duarte}
\author{S. Mukim}
\affiliation{School of Physics, Trinity College Dublin, Dublin 2, Ireland}
\author{A. Molina-S\'{a}nchez}
\affiliation{Institute of Materials Science (ICMUV), University of Valencia, Catedr\'{a}tico Beltr\'{a}n 2, E-46980, Valencia, Spain}
\author{ Tatiana G. Rappoport}
\affiliation{Instituto de Telecomunicações, Instituto Superior Técnico, University
of Lisbon, Avenida Rovisco Pais 1, Lisboa, 1049001 Portugal}
\affiliation{Instituto de Física, Universidade Federal do Rio de Janeiro, Caixa
Postal 68528, 21941-972 Rio de Janeiro RJ, Brazil}
\author{M. S. Ferreira}
\affiliation{School of Physics, Trinity College Dublin, Dublin 2, Ireland}
\affiliation{Centre for Research on Adaptive Nanostructures and Nanodevices (CRANN) \& Advanced Materials and Bioengineering Research (AMBER) Centre, Trinity College Dublin, Dublin 2, Ireland}
\begin{abstract}
\noindent
To calculate the conductivity of a material having full knowledge of its composition is a reasonably simple task. To do the same in reverse, i.e., to find information about the composition of a device from its conductivity response alone, is very challenging and even more so in the presence of disorder. An inversion methodology capable of  decoding the information contained in the conductivity response of disordered structures has been recently proposed but despite claims of generality and robustness, the method has only been used with 2D systems possessing relatively simple electronic structures. Here we put these claims to the test and generalise the inversion method to the case of monolayer MoS$_2$, a material whose electronic structure is far more complex and elaborate. Starting from the spectral function that describes the DC conductivity of a disordered sample of a single layered MoS$_2$ containing a small concentration of randomly dispersed vacancies, we are able to invert the signal and find the exact composition of defects with an impressive degree of accuracy. Remarkably, equally accurate results are obtained with the optical conductivity. This is indicative of a methodology that is indeed suitable to extract composition information from different 2D materials, regardless of their electronic structure complexity. Calculated conductivity results were used as a proxy for their experimental counterpart and were obtained with an efficient quantum transport code (KITE) based on a real-space multi-orbital tight-binding model with parameters generated by density functional theory. 

\end{abstract}


\maketitle


Transition Metal Dichalcogenides (TMDs) research has been in focus during the last decade due to the remarkable electronic and mechanical properties these transition metals display~\cite{Yuan, choi2017recent, chhowalla2015two, duan2015two, akinwande2014two}. Besides their applications in valleytronics~\cite{shan2013spin} and indirect to direct bandgap transitions ~\cite{wang2012electronics}, these are 2D materials that promise to fill the void left by graphene for possessing bandgaps with sizes that are far more useful for applications. For example, many TMD gaps are of the same order of magnitude as sunlight or infra-red photon energy~\cite{xia2014two}, which makes them particularly suitable for energy harvesting and photovoltaic functionalities.

Despite being technologically promising, the fabrication of these materials on a large scale is not so straightforward. Despite recent advances in 2D materials manufacturing technology~\cite{wang2018mass, zhang2020mass}, there is still a lack of atomic thickness precision to produce them on an industrial scale~\cite{yang2019electronic}. Furthermore, during the manufacturing of TMD mono-layers, some degree of disorder is inevitable and certain to affect the otherwise perfect crystalline structure of these layered materials. Due to the type of atomic bonding in monolayer TMDs, structural disorder in the form of vacancies of the chalcogens is one of the most likely to occur~\cite{li2017atomic}. Normally seen as detrimental to the properties of a crystalline material, disorder can also unlock new properties~\cite{Lin2016,schuler2019large} such as local magnetic moment~\cite{zhou2013electronic, yang2019electronic}, as well as be used for converting the compound into p- or n-type semiconductors.  It is thus paramount to distill as much information as possible about the nature and the level of disorder present in these materials to assess how that impacts their optoelectronic properties. 

\begin{figure}[h!]
	\includegraphics[width=\columnwidth]{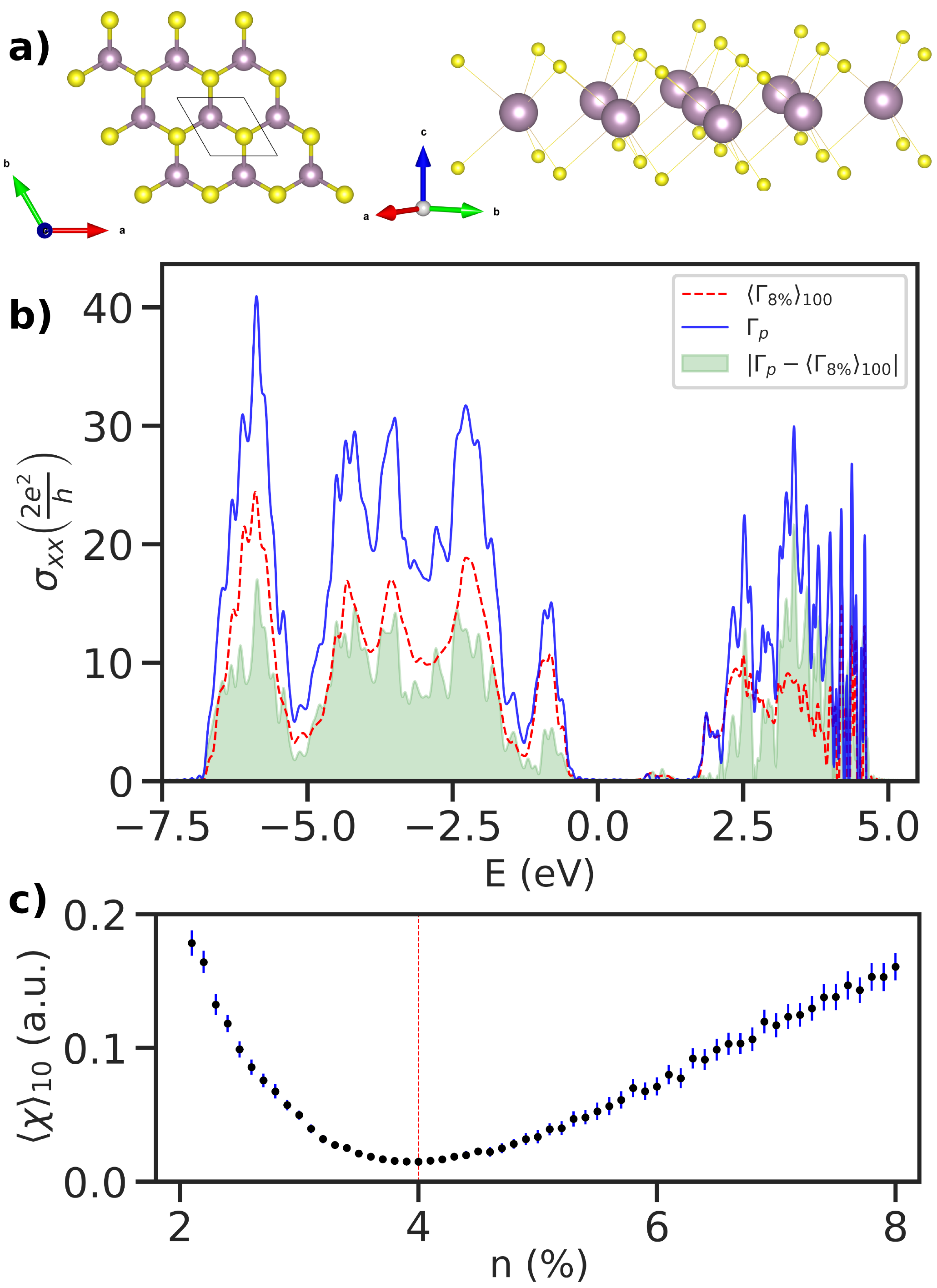}
    \caption{\textbf{a)} Top-down view and standard orientation of crystal shape for MoS$_{2}$ in 1H structural phase (purple sphere = Mo, yellow sphere = S). \textbf{b)} Longitudinal conductivity at temperature $T=0$K comparison between 8\% concentration CA and parent configuration. \textbf{c)} Misfit function $\chi(n)$ in arbitrary units averaged over 10 distinct parent configurations with minimum at 4\%. For the sake of comparison, the vertical (red) line indicates the real concentration $n_p$ of the parent configuration.}
\label{comp}
\end{figure}

While numerous characterisation tools can identify and quantify the type of disorder and dopants present in TMDs~\cite{zhang2015phonon, vancso2016intrinsic, wang2015physical}, some require a lot more effort and machinery than others. A recently developed methodology~\cite{shardul2020} that claims to have a simple inversion procedure to extract compositional information about disordered structures appears as a suitable candidate to quantify the number of defects in TMDs. In particular, it uses the energy-dependent conductivity as the only input based on which the inversion takes place. Claims of generality and robustness suggest that said inversion tool is applicable to a wide range of 2D materials, although evidence was only given with graphene, for which the electronic structure is rather simple. In this article, we put this method and its generality claim to the test by considering one specific type of TMD, namely MoS$_2$ (see Fig.\ref{comp}a), containing a small concentration of vacancies. We employ a multi-orbital tight-binding model
with parameters obtained from \textit{ab initio} calculations~\cite{Pizzi2019} to describe a far more complex electronic structure than the one for graphene. Starting from the seemingly noisy spectral conductivity we obtain the number of vacancies with an impressive level of accuracy. Furthermore, we have also adapted and extended the original ideas of the aforementioned inversion technique to work with optical signals instead of longitudinal conductances. Remarkably, the same level of accuracy, if not higher, was achieved with the optical conductivity, a different starting point that gives equally good results.

The nature of inverse problems in science is in obtaining from a set of observations the causal factors that generated them in the first place. Here we use the spectral (longitudinal and optical) conductivity as our starting point. These are quantities normally obtained by standard experimental setups of a two-terminal device but may be also calculated once the Hamiltonian is fully specified. In this manuscript we shall use the latter as a proxy for the former, i.e., calculated conductivity spectra will be used to represent their experimental equivalent. The obvious advantage of using calculated functions is that we can refer back to the disordered-system Hamiltonian that generated them in the first place, making it possible to assess the success of the inversion procedure.

To obtain the real-space hamiltonian for MoS$_2$, we have performed density-functional theory (DFT) calculations, within the local-density approximation (LDA), using the Quantum Espresso numerical packages~\cite{Giannozzi2009}. We have employed norm-conserving and fully relativistic pseudopotentials~\cite{Hamann2013,VanSetten2018}, and generated a basis of atomic orbitals using Wannier90~\cite{Pizzi2019} for MoS$_2$ that includes the $s$, $p_x$, $p_y$ and $p_z$ orbitals of the chalcogen  atoms and the five $d$ orbitals of the Mo (see exact composition of states at $K$ in Ref.~\cite{MolinaSanchez2015}). The  tight-binding Hamiltonian obtained with Wannier90 is exported to KITE with PythTB scripts~\cite{vanderbilt_2018}. Its band structure and density of states is consistent with DFT calculations~\cite{canonico2020orbital} (see SM) and multi-orbital tight-binding models~\cite{Cappelluti2013, Yuan2014, Ridolfi2015, Fang2015}.
Conductivity calculations were carried out using KITE~\cite{kite}, an efficient Quantum Transport (QT) software for real-space tight-binding simulations, capable of handling an extremely large number of atomic orbitals. It is based on Chebyshev polynomial expansions~\cite{weisse2006kernel,fan2018linear} and provides excellent accuracy levels without being too demanding in terms of computational resources. Furthermore, its real space formalism is ideally suited to study disorder effects on materials. Spectral properties such as Density of States (DOS)~\cite{Garcia14}, DC conductivity~\cite{garcia2015real,ferreira2015critical,Canonico2018,canonico2020orbital}, and optical conductivity~\cite{Joao2019} are some of the combined outputs of these calculations and can be easily tested against a variable degree of structural disorder. It is thus possible to create a collection of spectral quantities results for different configurations and concentrations of disorder. 

In practice, to obtain the spectral response from real disordered compounds such as the longitudinal conductivity is rather straightforward with KITE, but to find out the actual impurity concentration of the exact configuration that originated one specific spectral function is far from simple. A naive approach would be to run over the possible combinations of disorders in the hope of eventually finding the exact parent configuration that generated the input function in the first place. However, the enormity of cases required to guarantee that the parent configurations are found makes this approach impracticable. Rather than carrying out an extensive "blind" search, the inversion methodology~\cite{shardul2020} makes use of configurationally averaged signals combined with the ergodic assumption that averaging over energy (or frequency) is equivalent to considering different configurations of disorder. In reality, this corresponds to a rapid expansion of the number of configurations being compared against the parent signal.

In mathematical terms, this is captured by a functional $\chi(n)$ defined as
\begin{equation}
\chi(n) = \frac{1}{\varepsilon_{+}-\varepsilon_{-}}\int_{\varepsilon_{-}}^{\varepsilon_{+}} dE \, \left[\Gamma(E)- \braket{\Gamma(E,n)} \right]^{2} \, ,
\label{chi}
\end{equation}
hereafter referred to as the misfit function.

In Eq.(\ref{chi}), $\Gamma(E)$ is the input spectral function of the probing system with unknown disorder concentration and $\braket{\Gamma(E,n)}$ is the Configuration Average (CA) of that spectral quantity defined as
\begin{equation}
\braket{\Gamma(E,n)} = \frac{1}{N_c}\sum_{m=1}^{N_c} \Gamma_m(E,n)\,\,.
\label{CA-Gamma}
\end{equation}
The index $m$ labels the different disordered configurations used in the averaging whereas $N_c$ accounts for the total number of them. As we shall see, modest values of $N_c$ are sufficient to generate excellent inversion accuracy. Note that the input function $\Gamma(E)$ is a function of energy only, whereas its CA counterpart also depends on the impurity concentration $n$. The integration limits are arbitrary energy values and may span a variable fraction of the band structure. The misfit function $\chi(n)$ is easily interpreted as a quantity that measures the deviation between the input transmission of the parent configuration $\Gamma(E)$ and its CA counterpart $\langle \Gamma(E,n) \rangle$. On calculating the quantity $\chi(n)$ for different concentrations one gets a function that is expected to display a distinctive minimum at the real value of $n$ contained in the parent configuration. Furthermore, with a smart interpolation scheme based on a machine-learning strategy we are able to carry out the averaging of Eq.(\ref{CA-Gamma}) with a fine mesh for $n$ (see SM) \cite{machinelearning_modelling,machine_learning_interpolation}. Although not an essential step in the calculation, this improves the resolution of $\chi(n)$ and enhances the accuracy of the inversion process with only small computational costs.

The equivalence between increasing the number of configurations in the CA part of the calculation and increasing the number of integrated energy points enables us to replace having a configuration average containing large amounts of functions with one with fewer functions that carry a higher energy point number - be it by increasing the energy point density or by spanning a wider range of the spectrum. This makes KITE particularly suitable because when using the Kernel Polynomial Method (KPM), KITE's output is evaluated at as many energy points as desired without changing the computational demands since it provides the function for the whole energy spectrum~\cite{weisse2006kernel,kite}. Therefore, the number of configurations $N_c$ needed in the CA part of the calculation may be reduced without impacting the inversion accuracy. 

With that in mind, we obtain numerous results for the longitudinal DC conductivity and optical conductivity spectral function, both calculated through KITE. All calculations were carried out with $64\times64$ unit cells comprising a total of 13 orbitals per unit cell (4 orbitals for each S atom and 5 orbitals for the Mo atom) with a finite concentration of disorderly distributed chalcogen vacancies, which in the case of the MoS$_{2}$  are sulfur (S) vacancies. 
For each function, 256 moments of Chebyshev polynomials expansion were used for 9 different random vectors trace calculations. Our results for clean samples are consistent with previous theoretical analyses based on multi-orbital tight-binding hamiltonians~\cite{Roldan2016,Yuan2014,Li2012}. Despite working with a small system size, few random vectors and few polynomial moments, the inherent noise of the KPM approach should be averaged out when calculating the CA. The same can be said about the quantity $\braket{\chi}$ regarding the parent configurations noise. This feature of the calculation appears to contribute to the efficiency of the method, allowing for low cost calculations without loss in precision.

Notice that the calculations were carried out at temperature $T=0$K. Even tough this is not a typical experimental temperature, the method still works in the case of relevant temperatures, as discussed in the Reference \cite{shardul2020}. The effect of typical experimental temperatures are much smaller than the correlation length of our conductivity energy integrations and therefore meaningful information can still be extracted from the Eq.\ref{chi} at higher temperatures. In fact, one could use the same low temperature CA calculations for obtaining misfit functions from room temperature input functions provided its correlation length stays greater then the temperature effect.

One specific set of vacancy positions is arbitrarily chosen to be the parent configuration, which in this case contains $n_p=4\%$ of vacancies. Note that this information is not used in any part of the subsequent calculation, except for testing the inversion success at the end. Fig.\ref{comp}b captures the essence of the inversion approach by showing the conductivity spectrum of the parent configuration (blue solid line) together with a CA calculation with $N_c=100$ configurations for $n=8\%$ of vacancies (red dashed line), which in this case enters as a simple guess. There is quite a discrepancy between the two curves, suggesting that $8\%$ is not the real vacancy concentration in the configuration that generated $\Gamma(E)$. In fact, the green-filled curve corresponds to the absolute value of the deviation between the red and blue curves. According to Eq.(\ref{chi}), a simple integration of that deviation curve leads to a direct calculation of $\chi$, in this case for the specific case of $n=8\%$.

Repeating this process for different concentrations $n$, we get the misfit function $\chi(n)$, seen in Fig.\ref{comp}c. The distinctive minimum at $n=4\%$ coincides with the exact vacancy concentration $n_p$ contained in the parent configuration and is yet another evidence of a successful inversion. The same procedure was repeated 10 times with distinct parent configurations and each time the inverted concentration was compared with the real value, leading to a statistical distribution of the misfit function $\braket{\chi(n)}_{10}$. This appears in the figure in the form of error bars.

\begin{figure}[h!]
  \begin{center}
	\includegraphics[width=\columnwidth]{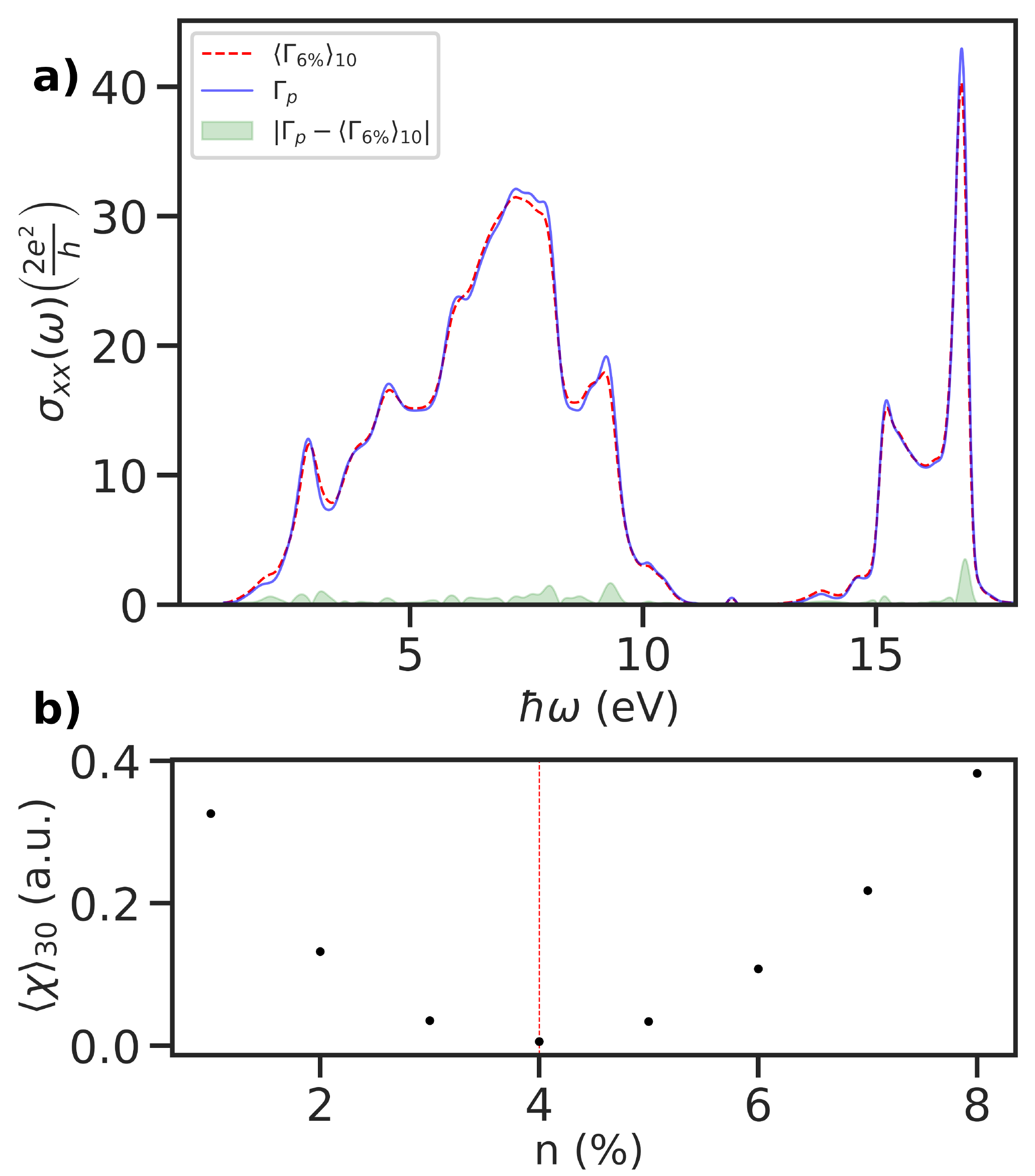}
  \end{center}
\caption{\textbf{a)} Optical conductivity at temperature $T=0$K comparison between 6\% concentration CA and parent configuration. \textbf{b)} Misfit function $\chi(n)$ in arbitrary units averaged over 30 distinct parent configurations with minimum at 4\%. The error bars are smaller than the point markers. For the sake of comparison, the vertical (red) line indicates the real concentration $n_p$ of the parent configuration. 
}
\label{data}
\end{figure}

One parent configuration is enough for performing the inversion. However, to assess the accuracy of the inversion we must repeat the process several times with different configurations in order to obtain statistical information for a given concentration. The higher the number of parent configurations used, the higher is the statistical significance of our data.

Regarding previously made claims of generality and robustness~\cite{shardul2020}, the inversion must be tested with an input function other than the electrical conductivity. Although more computationally demanding, the optical conductivity is a suitable candidate. Following similar steps to the ones taken earlier, the spectral optical conductivity for a parent configuration as well as for numerous other configurations were calculated. Shown in Fig.\ref{data}a are the parent signal (blue solid line) together with its CA counterpart (red dashed line) for an arbitrary concentration value of $n=6\%$. Once again, the concentration-dependent misfit function $\chi(n)$ is generated when the deviation between the two curves is integrated over energy. Fig.\ref{data}b depicts $\chi(n)$ for concentration values between $1\%$ and $8\%$. Reassuringly, even though the parent signal is of a different nature and looks very different to the one used in Fig.\ref{comp}b, the inversion finds exactly the same answer with a distinctive minimum at $n=4\%$.

Remarkably, the error bars in this case are of comparable sizes to the DC conductivity inversion even in cases where the CA calculation involved considerably lower $N_c$ values.
We speculate that this is a peculiarity of the electronic structure of MoS$_{2}$ which may forbid the propagation of certain wave vectors and in turn limit the probing of the disordered environment, something that can then only be achieved with higher numbers of $N_c$. There are no such restrictions for the propagation of electromagnetic waves and therefore the optical conductivity requires far fewer configurations in order to fully capture the disordered environment of a given parent signal.
The optical conductivity calculation for each energy point takes as input the energy values in a range of the order of $\hbar\omega$ around the calculated point. In other words, the optical conductivity calculations involve a larger volume of states which, according to the ergodic principle is equivalent to a larger number of iterations. This ultimately translates into a grater accuracy for the misfit function calculation.

\begin{figure}[ht!]
  \begin{center}
	\includegraphics[width=\columnwidth]{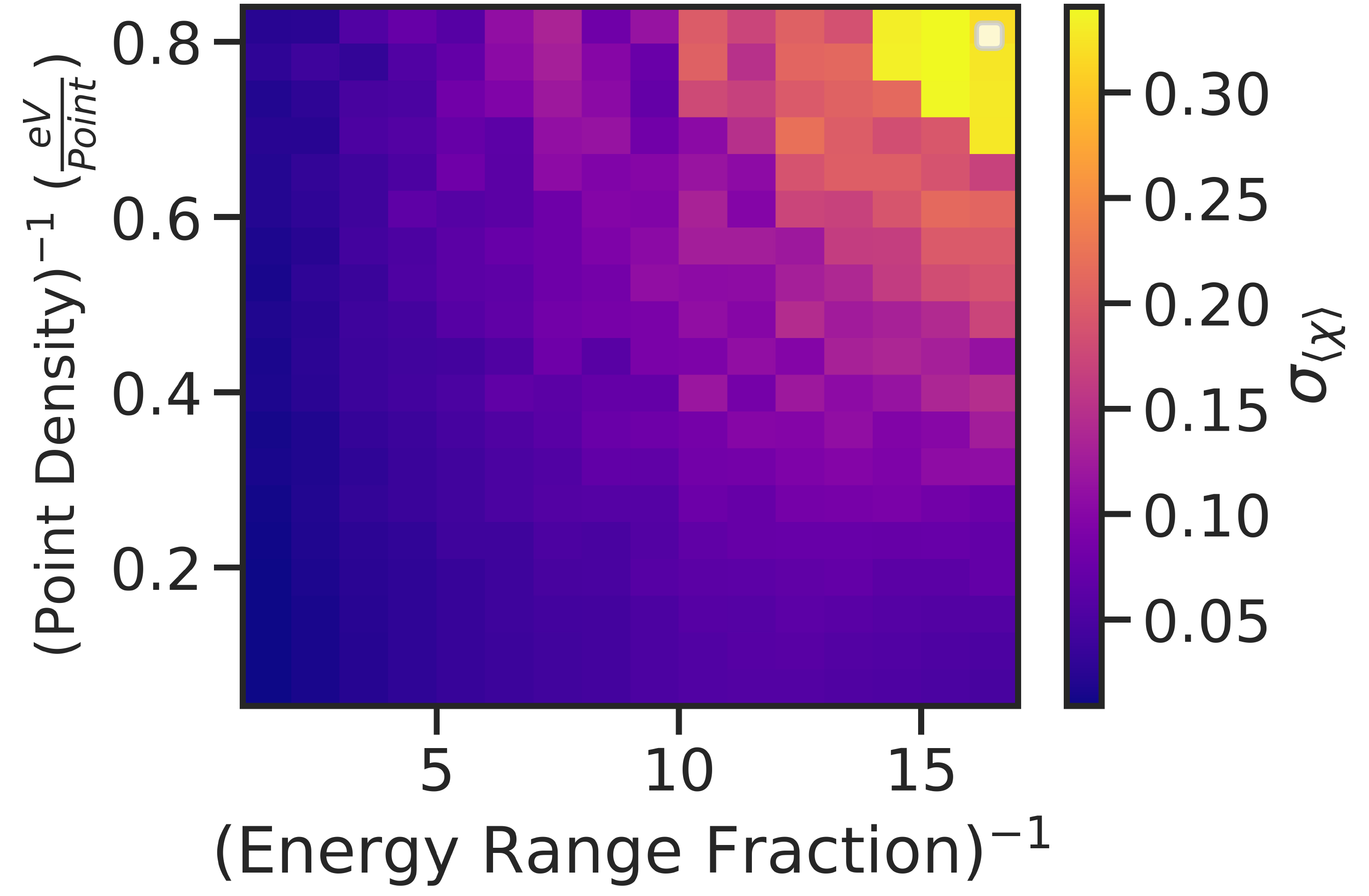}
  \end{center}
\caption{Behavior of the mean standard deviation as a function of the point density and the energy range covered. Statistical distribution for 50 different parent configurations with CA of 50 configurations $\sigma_{ \braket{\chi_{50}}_{50}}$.}
\label{optdata}
\end{figure}

One question worth asking is how essential the integral of Eq.(\ref{chi}) is. On the one hand, the integration plays a significant part in identifying the real impurity concentration because fluctuations of a single sample versus energy (frequency) are equivalent to sample-to-sample fluctuations at fixed energy (frequency). Therefore, the integration is similar to vastly augmenting the number of disordered configurations taken into account. On the other, the integration itself may be replaced with a discrete sum without affecting the accuracy of the inversion procedure. By reducing the number of energy (frequency) points involved in generating the misfit function, there should be a threshold below which the inversion accuracy drops quite significantly. This is indeed the case and is depicted in Fig.\ref{optdata} as a color plot of the standard deviation associated with the inversion accuracy as a function of the reciprocal of two quantities, namely the energy range fraction and the integrated point density. The energy-range fraction refers to the proportion of the electronic band structure taken into account in the inversion calculation. The point density refers to how dense an energy mesh was used in the integration of Eq.(\ref{chi}). The plot unmistakably shows that the higher the density of points or the spectral range used, the lower the standard deviation we get from the same number of configurations in the CA calculation. Furthermore, the bright spot on the top right of the plot indicates that the inversion accuracy drops if both the point density and the energy range fraction are reasonably low. This may be interpreted as an answer to the question posed at the beginning of this paragraph, because for a sufficiently wide energy range fraction, the integration point density may be lowered to fairly small values without impacting the inversion accuracy. That is another way of saying that in this case the integration can be safely treated a sum.

It is worth emphasizing that the accuracy analysis of Fig.\ref{optdata} not only validates the ergodic assumption behind the inversion methodology but also demonstrates the robustness and efficiency of this inversion method. In both cases, it was possible to pinpoint the exact concentration of defects contained in the probing system with great accuracy even when decreasing the amount of data used for the configuration averages. This behaviour was seen for parent configurations with defect concentrations of up to 8\%. Finally, it is worth pointing out that this study also suggests that the method applicability can be extended to any type of spectral function besides the DC and optical conductivities. Combined with the KPM method, it can determine the disorder concentration of realistic systems with a reasonably low computational cost and points to possible on-demand inversion of experimental spectral signals. On attempting such inversion from an experimental setup, it is important to consider the possible relevant disorder types on the system in hand, such as different scatterers, and calculate the CA accordingly. The method can then easily be extended for the case of multiple disorder types as seen in Reference \cite{shardul2020}.

In summary, we have shown that information about the composition of disordered structures can be extracted from seemingly noisy conductivity response functions through an efficient inversion methodology. The method is not material-specific and can be used with a variety of 2D materials, from simple electronic structure such as graphene~\cite{shardul2020} to more complex materials such as the case shown here with MoS$_2$. Remarkably, the inversion works not only with the DC conductivity, as originally proposed, but is equally accurate with a different spectral function such as the optical conductivity. Such a robust and versatile methodology suggests that structural and compositional information about quantum disordered devices can be extracted from other physical signals beyond the DC and optical conductivities.


\typeout{}
\bibliography{mybib}{}
\bibliographystyle{unsrt}

\section{Aknowlegments}
 \begin{acknowledgements} \noindent TGR acknowledges funding from Fundação para a Ciência e a Tecnologia and Instituto de Telecomunicações - grant number UID/50008/2020 in the framework of the project Sym-Break. A. M.-S. acknowledges the Ram\'on y Cajal programme (grant RYC2018-024024-I; MINECO, Spain). Ab initio simulations were performed performed on the Tirant III cluster of the Servei d‘Informática of the University of Valencia.
\end{acknowledgements}

\section{Author Contributions}
\noindent FRD cowrote the paper and along with SM conducted all the spectral simulations and data analysis. AM made the ab-initio calculations, data from which was used to obtain the simulations parameters. TGR led the quantum transport implementation. MSF led the research and cowrote the paper. All authors discussed and commented on the manuscript and on the results.

\section{Aditional Information}
\noindent\textbf{Competing Interests.}
The authors declare no Competing Financial or Non-Financial Interests.

\noindent \textbf{KITE.}
KITE has an open-access repository at \url{https://zenodo.org/record/3245011} and \url{https://quantum-kite.com/}.

\noindent \textbf{Data Availability.}
Data available on request from the authors.


\end{document}


\title{SM:Decoding the DC and optical conductivities of disordered MoS$_2$ films: an inverse problem} 
\author{F. R. Duarte}
\author{S. Mukim}
\affiliation{School of Physics, Trinity College Dublin, Dublin 2, Ireland}
\author{A. Molina-S\'{a}nchez}
\affiliation{Institute of Materials Science (ICMUV), University of Valencia, Catedr\'{a}tico Beltr\'{a}n 2, E-46980, Valencia, Spain}
\author{ Tatiana G. Rappoport}
\affiliation{Instituto de Telecomunicações, Instituto Superior Técnico, University
of Lisbon, Avenida Rovisco Pais 1, Lisboa, 1049001 Portugal}
\affiliation{Instituto de Física, Universidade Federal do Rio de Janeiro, Caixa
Postal 68528, 21941-972 Rio de Janeiro RJ, Brazil}
\author{M. S. Ferreira}
\affiliation{School of Physics, Trinity College Dublin, Dublin 2, Ireland}
\affiliation{Centre for Research on Adaptive Nanostructures and Nanodevices (CRANN) \& Advanced Materials and Bioengineering Research (AMBER) Centre, Trinity College Dublin, Dublin 2, Ireland}

\maketitle


\section{Interpolation Method for CA}
To demonstrate the agreement between minimizing concentration of $\chi$ and the actual disorder concentration the configurationally averaged (CA) conductance is expanded to linear order in $n$ as \cite{datta_1995}
\begin{equation}
\langle \Gamma(E,n) \rangle = \Gamma_0(E) - \beta(E) \, n,
\label{CA-approx}
\end{equation}
where $\beta(E)$ is the derivative of $\langle \Gamma(E,n) \rangle$ with respect to $n$ evaluated at $n=0$. 

\begin{figure}[!h]
    \includegraphics[width=\columnwidth]{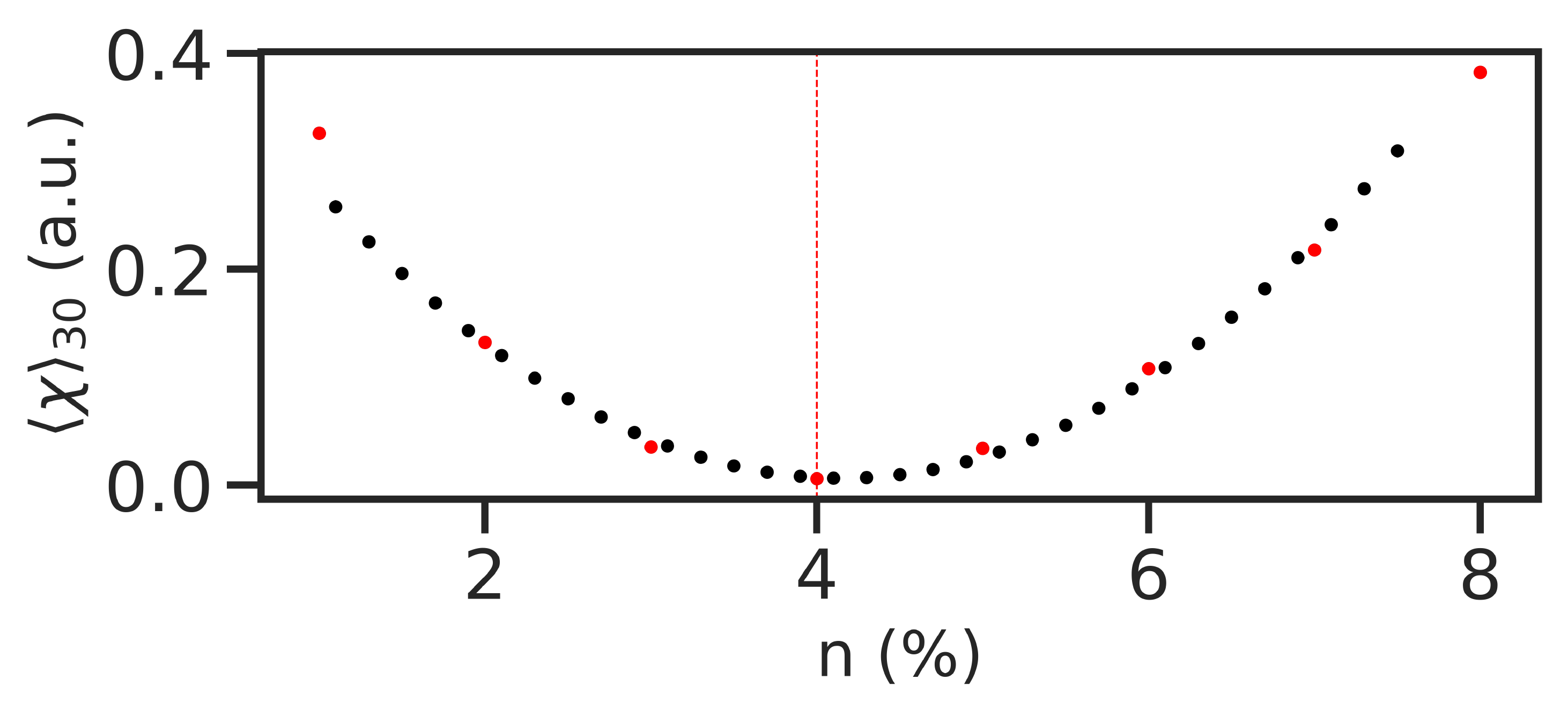}
    \caption{Misfit function for the machine learning optical conductivity interpolated data (black) and for the numerical data (red). The model still predicts the minimum correctly while providing more resolution around it, allowing for fine probing different parent configuration concentrations with lower computational cost.}
    \label{interpol}
\end{figure}

Satisfactory accuracy levels  of the inversion method relies on carrying out the CA part of the calculation with a good resolution in terms of $n$ but the computational costs of doing so with a large number of concentration points by brute force is substantial. The approximation in Eq.\ref{CA-approx} provides a doorway to generate CA for more values of $n$ using interpolation supported by machine learning. Fig.~\ref{interpol} shows a misfit function generated using numerical (red) and interpolated (black) data. This model studies the CA data (CA corresponding to red points) to generate CA spectrum for different values of $n$ (corresponding to black points). Eq.\ref{CA-approx} is valid for a small range of concentration but it starts to move away from the linear nature as disorder concentration increases. This results in failure to understand the caveats of CA spectrum.

\begin{figure}
    \includegraphics[width=\columnwidth]{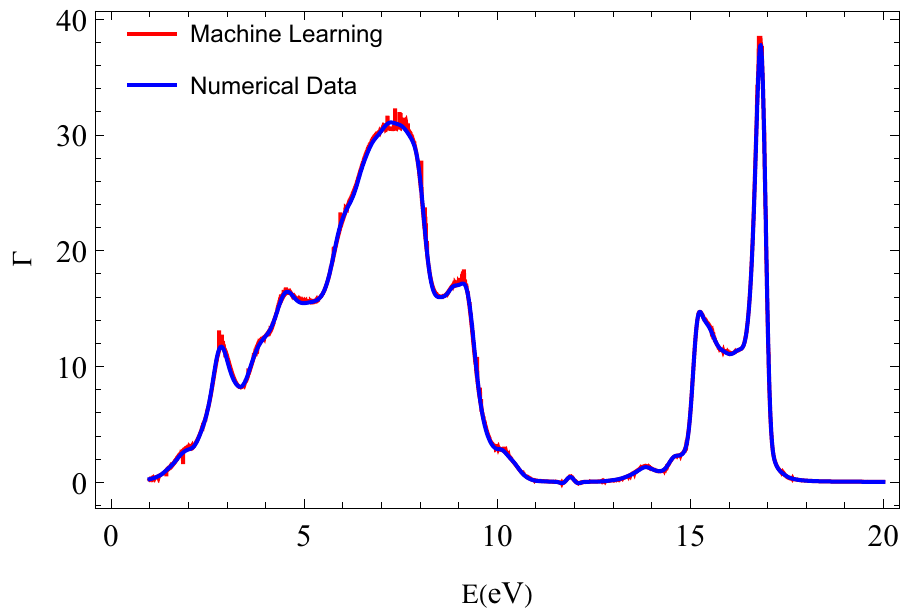}
    \caption{Configurational average corresponding to 8\% disorder randomly distributed in the device represented in Blue. In Red, ML optical conductivity shows very good prediction with the numerical result. }
    \label{machinelearning}
\end{figure}

Machine Learning (ML) becomes a very handy tool in order to generate CA conductance spectrum. It computes CA spectrum without losing details of the conductance quantum signatures and at a low computational costs. Fig.~\ref{machinelearning} shows very good prediction of CA optical spectrum corresponding to 8\% disorder with numerically calculated optical spectrum in blue. 7 datasets corresponding to different concentrations of numerically calculated CA spectrum were used to train the model. Same model is used to generate rest of the CA points shown in black in Fig.~\ref{interpol}.

Wolfram Mathematica provides predefined ML tools called Predict and Classify to define a model. Both tools can generate the output corresponding to concentration and energy in this case. Model uses numerically computed CA spectrums as the training examples. Spectrum provides large number energy points to train the model.
It is worth pointing out that while more sophisticated ML techniques are available to model quantum transport \cite{machine_learning_interpolation,machinelearning_modelling}, this is by no means an essential ingredient of this inversion method but one that can speed up the inversion and improve its accuracy.

\section{MoS$_2$ Band Structure and conductivity for clean systems}
\begin{figure}
    \includegraphics[width=\columnwidth]{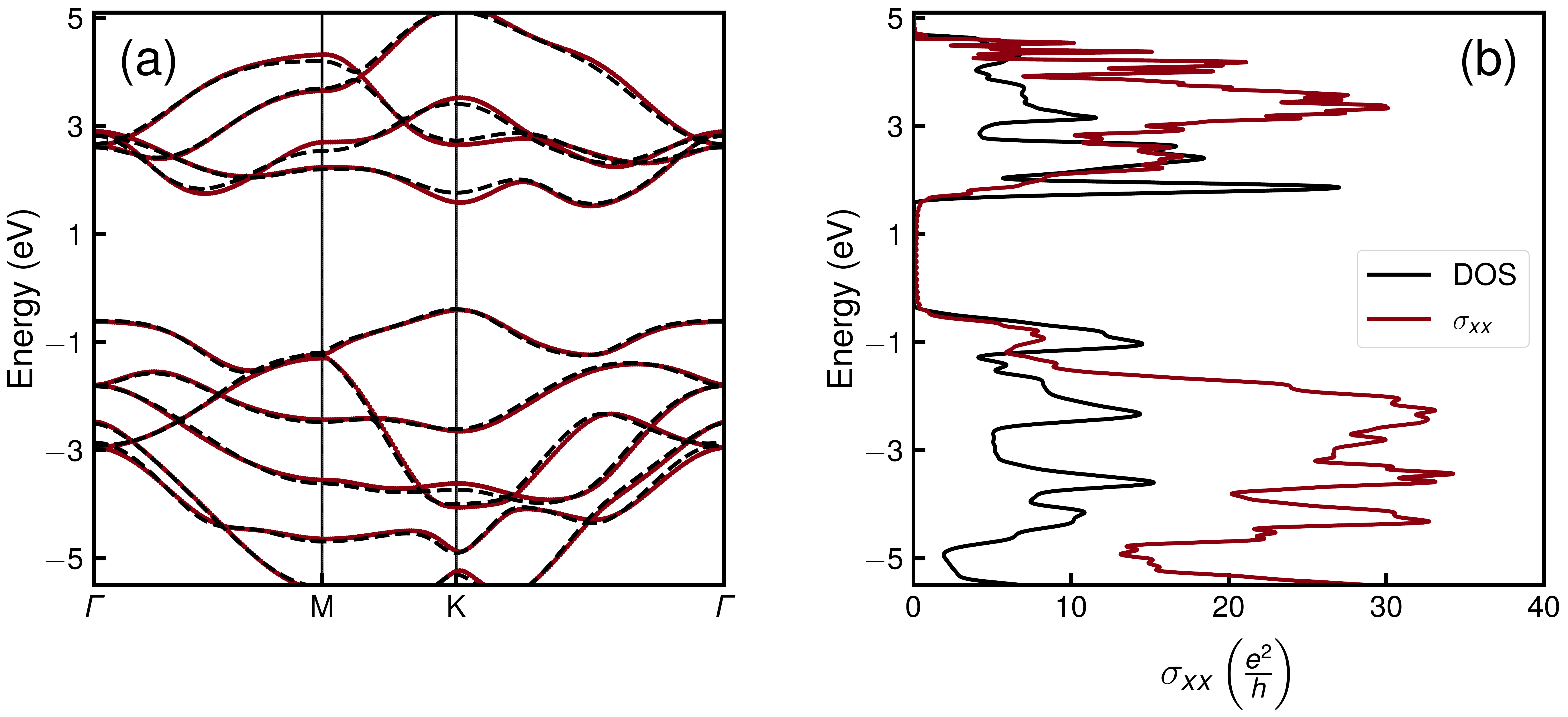}
    \caption{(a)Comparison between the band structures of a MoS$_2$ monolayer calculated without SOC employing DFT (solid line) and the effective tight-binding model (dashed line). (b) Density of states (black) and conductivity (blue) of  MoS$_2$ calculated with KPM and the effective tight-binding model.}
    \label{DOS}
\end{figure}

\begin{figure}[ht!]
    \begin{center}
        \includegraphics[width=\columnwidth]{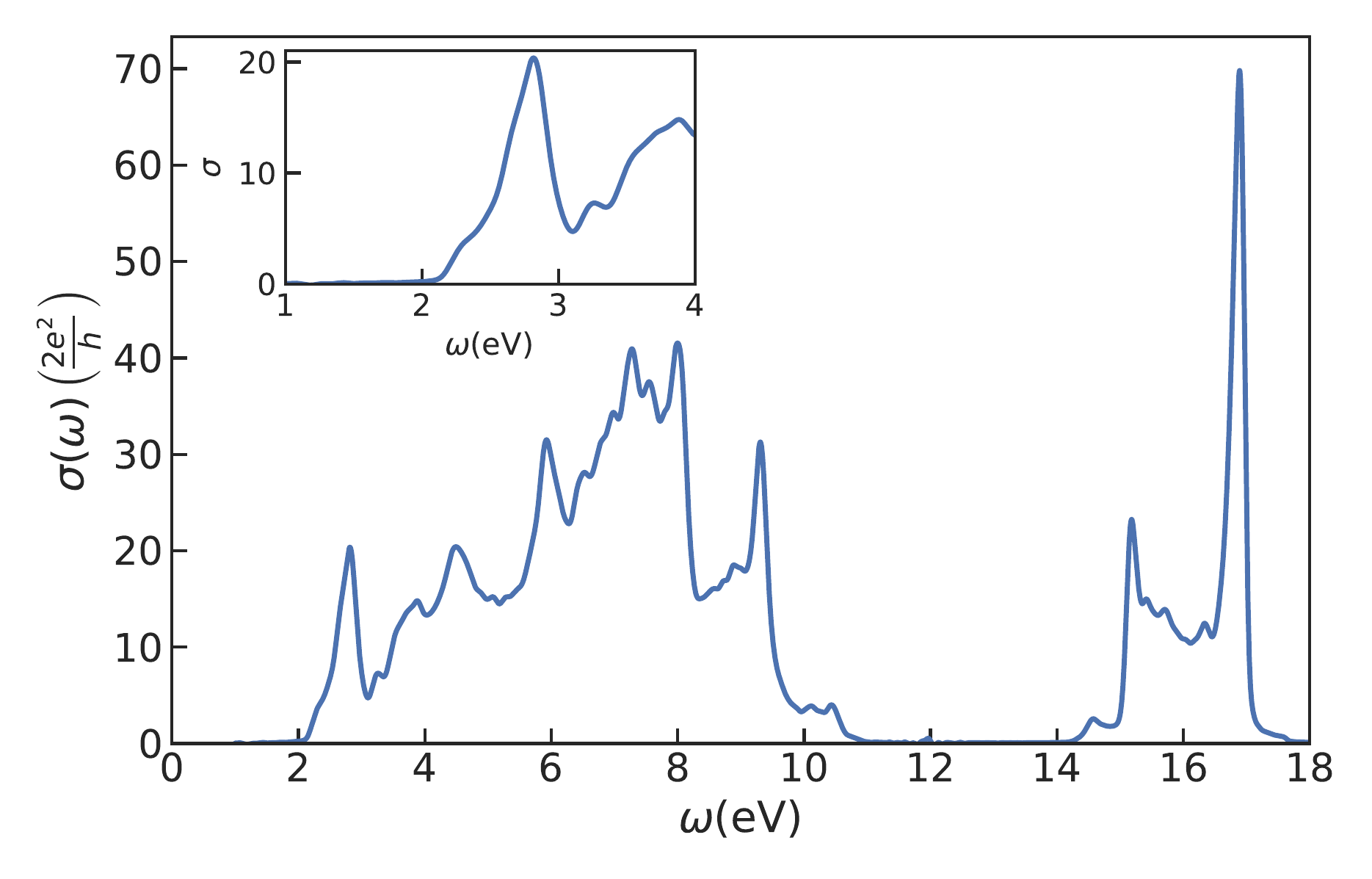}
    \end{center}
    \caption{Optical conductivity of MoS$_2$ calculated with KPM and the effective tight-binding model with Fermi energy at 0 eV.}
    \label{optcond}
\end{figure}

 Fig. \ref{DOS}(a) shows the comparison
 between the band structure calculations for MoS$_2$ obtained by
 DFT and by the effective tight-binding Hamiltonian taking
 into account 13 orbitals per unit cell. We have considered all hopping integrals with energies higher than 0.0125 of the maximum hopping value. 
Fig. \ref{DOS}(b) presents the density of states and the conductivity without disorder, both calculated with the kernel polynomial method for the effective tight-binding Hamiltonian. Fig. \ref{optcond} show the optical conductivity pristine case.


\typeout{}